# Enhancing Byzantine fault tolerance using MD5 checksum and delay variation in Cloud services

C Sathya , S Agilan , A G Aruna
Coimbatore Institute of Technology,
Coimbatore-14.Tamil Nadu, India.

**Abstract**
Cloud computing is an attempt to stretch the operation boundaries dynamically on-demand which is not the case with traditional distributed system. This unprecedented nature of cloud though offers various benefits it also has a downside when it comes to fault tolerance since most of the faults experienced by the cloud are unpredictable and thus Byzantine in nature. Byzantine faults in cloud computing is a serious problem because of the fault propagation possibility exists which makes the fault to propagate surreptitiously through consecutive virtual components still it spreads throughout the entire cloud system. Consequently the real-world problems faced by the cloud computing management are beyond typical human narratives. However if a virtual system is not effectively designed to tolerate Byzantine faults, it could lead to a faultily executed mission rather than a cloud crash. The cloud could recover from the crash but it could not recover from the loss of credibility. Moreover no amount of replication or fault handling measures can be helpful in facing a Byzantine fault unless the virtual system is designed to detect, tolerate and eliminate such faults. However research efforts that are made to address Byzantine faults have not provided convincing solutions vastly due to their limited capabilities in detecting the Byzantine faults. As a result, in this paper the Cloud system is modeled as a discrete system to determine the virtual system behavior at varying time intervals. A delay variation variable as a measure of deviation for the expected processing delay associated with the virtual nodes takes values from the set of P {low, normal, high, extreme}. Similarly, a check sum error variable which is even computed for intra nodes that have no attachment to TCP/IP stack takes values from the set of P {no error, error}. These conditions are then represented by the occurrence of faulty events that cause specific component mode transition from *fail safe* to *fail-stop or byzantine prone*. The correlation analysis is performed and the result shows that the proposed system show considerable improvement in detecting the Byzantine faults.

## 1. Introduction

Cloud computing is the Internet based computing which provides elastic services and allot resources on-demand. Cloud computing involves using data center servers and software networks to dynamically allocate resources and applications on-the-fly for remote end users. Since the arrival of distributed computing various other concepts based upon collective resource utilization using Internet as back bone has culminated into flexible Cloud computing. It managed to offer solutions from expensive enterprise level such as private cloud to cost effective casual user level solutions such as public cloud *Chunye et al (2010)*. This helped the service providers to bursting the clouds i.e to configure the resources from both ends to meet their requirement thus the hybrid cloud emerges *Bhaskar et al (2009)*. For instance a company could vastly use public cloud services but to handle occasional bursts in activity it could summon the on-demand private cloud services in to its armory. Whereas for users who look for cost effective solutions can utilize the private clouds for certain sensitive component while hosting other components on the public cloud.

Moreover Cloud computing as a service-oriented architecture sees everything as service. Conversely the Cloud computing service providers offer services according to the three standard models they are Infrastructure as a Service (IaaS), Platform as a Service (PaaS), and Software as a Service (SaaS) *Bhaskar et al (2009)*. *Software as a Service (SaaS)* is the capability provided to the consumer to use the provider's applications running on a cloud infrastructure. The applications are made accessible from various client devices through either thin client interface or program interface. *Platform as a Service (PaaS)* provides cloud capability to the consumer to deploy consumer created or acquired applications created using programming languages, libraries, services, and tools supported by the provider. *Infrastructure as a Service (IaaS)* allows the consumer to provision processing, storage, networks, and other fundamental computing resources *Chunye et al (2010)*. However to establish the service oriented architecture the cloud service providers enable features such as Resource pooling, Rapid elasticity, Measured service, On-demand self-service, auto scaling, interoperability, multitenancy, flexibility in cost, space and time improvisation etc. This has led the cloud computing to see exponential growth. Due to its business success many cloud services emerges in to market which are usually large-scale and very complex to manage.

In all the offered services the consumer does not manage or control the underlying cloud infrastructure including network, servers, operating systems, storage, and applications. However for the customer, limited control over selecting and configuring them may be brokered in cloud Service level agreement (SLA). Consequently examining the cloud at its functional levels of services to make it a fault tolerant system becomes a more pressing research concern. The entire Cloud functionality revolves around virtualization. Virtualization is the process of creating virtual version of components such as computing hardware, storage devices, and computer network resources *Rohan et al (2012)*. The criticality behind virtualization is to host various virtual components on single real component to help utilizing resources to its maximum potential and maximum availability. The dynamically created virtual components are often termed as virtual machines for those who argue cloud is no longer a virtualization it is termed as nodes. Nevertheless using virtual machines for various mission-critical applications also creates several challenges at all levels of cloud service functionality. Key challenges involves but not limited to; developing critical virtualization solution to suit HPC needs *Chunye et al (2010)*, managing critical virtual machines and proper utilization of advanced capabilities enabled by virtualization for fault tolerance such as VM pause, VM checkpoint, and VM migration *Stephen et al (2010)*.

The challenges are often the result of unpredictable behavior of the nodes or of the transient links between them. Unlike usual software or a hardware which has an operational boundary whereas cloud computing is an attempt to stretch this operation boundaries. This is unprecedented and thus the faults the cloud experience mostly is unpredictable and undetectable therefore fatal in nature. Such fatal faults often have the potential to crash the nodes, to cause link disconnection and may also result in complete cloud shutdown. For example, LinkUp (MediaMax) a Cloud service provider went out of business after losing 45% of stored client data due to a single error *Christian et al (2009)*. Amazons supposedly fool-proof mission-critical EC2 cloud services crash leads to permanent loss of sensible data the story continues.

As a result, no matter how comprehensive the management solutions are, the real-world challenges faced by the cloud computing is expected to evolve and thus remains a challenge.

However if a virtual system is not effectively designed to tolerate such faults, it could lead to a defective execution or complete cloud crash *Kevin et al (2003)*. Research efforts made to address fault tolerance issues have not provided convincing solutions vastly due to their limited capabilities in detecting the faults. As a result, the research community is still thriving for necessary solutions to achieve promising reliability and dependability criterion even when byzantine faults occurs *Kevin et al (2003)*. The following section look into various aspects of the faults that can lead up to byzantine faults and the succeeding section will discuss the methods necessary to face the byzantine fault with minimal damage to availability and reliability criterion.

## 2. Existing System

### 2.1 Cloud Fault and Solutions a Critical review

The fault occurs due to the malfunction or deviation from expected course of action or due to the involvement of malicious element *Rafael et al (2006)*. Tolerance is the existence of methods to handle such faults to sustain the continuous functionality of the system.

### 2.1.1 Cloud Faults

Fault occurs due to reasons such as hardware failure, software bugs, operator error, network problems, security breaches, virtual component failure, virtual link failure etc. Faults occur in cloud can be classified into one of three categories:

*Transient faults* usually occur in the communicating channel between processing nodes either in intra or inter cloud nodes. If the node is connected to another cloud node through TCP/IP protocol then it is **inter node** *John et al (2016)*. Whereas the nodes which are part of the same system still connected to each other through a transient links are highly virtual in nature and does not follow the TCP/IP stack is termed as **intra node.** Transient faults occur once and mostly disappear due to the resilience in the working mechanism in case of inter nodes. However in case of intra nodes the transient faults are more persistent and often requires manual troubleshooting.

*Intermittent faults* occur at irregular intervals in a device or system that functions normally at other times. Intermittent fault occurring in intra node are much more complicated then the inter node faults and have the potential to evolve as byzantine fault. Detecting them becomes more challenging since it produces results even when the fault occurs. The real system can be equipped with various features such as alert raising mechanism, bypassing the error element, etc. However grooming the Virtual Machines to face such errors still remains a challenge due to the lack of detection mechanisms *Ravi et al (2013)*.

*Permanent faults* are the mostly occurring failure in Cloud intra nodes. It persists until the faulty virtual component is dealt with. It is different from traditional repair or replace scenario as exist in inter nodes. Since a single persistent node or transient failure can render the entire cloud mission a failure, because in most cases all the involved virtual components are trying to achieve a single mission through undertaking its various parallel modules *Kevin et al (2003)*.

Any of these faults can transpire either as *fail-silent* failure or as *Byzantine failure*. A fail-silent fault is predictable and detectable because the faulty units usually stop functioning and generate no output or produces bad output that clearly indicates failure *Ravi et al (2013)*. A Byzantine fault is unpredictable and undetectable since the faulty unit continues to function but produces incorrect outputs often appears as correct output. However Byzantine faults in cloud computing is a serious problem because of the fault propagation possibility exists which makes the fault to propagate through consecutive virtual components still it spreads throughout the entire system.

### 2.1.2 Fault Handling Prototype

Following is the successive models that are considered to handle faults from cloud design level before implementation to the management level after implementation.

*Fault avoidance* is a process of troubleshooting Cloud design and validation steps to ensure that the system avoids fault occurrences. It is very challenging in highly virtualized environment such as cloud because the paradigm of faults usually does not fall within the range of assumptions made and test cases derived for precursory fault evaluation.

*Fault tolerance* is the realization that the faults in the cloud system are inevitable so it is designed in such a way that the system compensates the faults and continue operation *Ifeanyi et al (2013)*. However there were scenarios when the fault happens it goes undetected thus results in complete system failure such as platform error. Fault tolerance can only be achieved in cloud systems if only the system have the capability to detect the faults before it propagate throughout the system.

*Fault removal* is the process of removing the fault through testing, debugging, and verification as well as replacing failed components. It can only happen after detecting the fault while it is encountered in the system which is becoming hard to realize due to the existence of Byzantine faults.

### 2.1.3 Fault handling Mechanisms

*Replication:* is the use of additional hardware, software and network resources to ensure duplication, it is to make sure a replica of a virtual component has been created to replace the failed virtual component in case of failure. *Replication mechanisms can be active or passive Ravi et al (2013).* The primary difference between synchronous (active) replication and asynchronous (passive) replication is the way in which data is written to the replica. Most synchronous replication is costly real-time approach because it writes data to primary storage and the replication storage simultaneously. Therefore the primary copy and the replica stay synchronized. In contrast asynchronous replication is cost effective and near real-time method which writes data to the primary storage first and then copy the data to the replication storage. Asynchronous replication is often scheduled at convenient timings.

*Checking and monitoring* the performance virtual machines is the key requirement for cloud service providers to meet the Cloud QoS requirements *Ravi et al (2013)*. The metrics used to monitor the cloud platform are not limited to CPU Percentage, Data In, Data Out, Disk Read Throughput, and Disk Write Throughput etc. These metrics enable the service provider with techniques necessary to detect the key failure and the subsequent reconfiguration.

*Checkpoint and Restart:* The system state is captured and saved based on pre-defined parameters such as for regular time intervals. When the system experience a failure, it is restored to the previously saved correct state using the latest checkpoint information, thus prevent the system from the complete restarts *Ifeanyi et al (2013)*.

The presented mechanisms come handy when handling any kind of faults, however customizing them to realize the full potential remains the challenge. Since quiet often the management team is puzzled with what to look for in case of subtle byzantine faults.

## 2.2 Literature review

**Martin et al (2006)** proposed a FaB Paxos based Byzantine consensus protocol to achieve consensus in two communication steps in common case. The cost for the common case two-step termination is high due to the higher number of acceptors as $5f + 1$ required. **Miguel et al (1999)** describes a state-machine replication algorithm to tolerate Byzantine faults in asynchronous systems. However reducing the resource replication still remains the challenge. **Yilei et al (2011)** presents a BFTCloud framework to work in voluntary resource cloud through selecting nodes based on their QoS performance. However the system is indefinite when facing byzantine faults which produce better QoS performance but gives inconspicuous output errors. **Jun et al (2010)** presents Read fault reduction and Write fault prediction methods to reduce page faults significantly. Further observation can be made to assess its capability to handle byzantine faults. **Rudiger et al (2012)** presents a CheapBFT system to tolerate byzantine fault with one active replica in normal-case operation thus requires only $f + 1$ active replica. Since it relies on intensive FPGA for authentication it can be time consuming so suitable only for homogeneous hpc systems. **Pedro et al (2011)** present an algorithm and prototype that tolerate byzantine faults in MapReduce. The presented algorithm can be further extended for larger cloud scenarios. **Pierre et al, (2013)** has proposed a Redundant Byzantine Fault Tolerance (RBFT) algorithm to run several instances of a BFT protocol in parallel, and to monitor their performance in order to detect a malicious primary. The parallel instances of BFT may be lavish in terms of performance. **Dominic et al, (2005)** identifies three goals as (1) to add fault tolerance without modifying existing system, (2) to minimize the time spent in executing non fault tolerant software, and (3) to minimize the time and space overhead needed to detect and recover from faults. ExtraVirt accomplish these goals by leveraging VM technology, sharing memory and I/O devices across replicas. **Binoy et al, (2003)** present two proactive resource allocation algorithms, RBA-FT and OBA-FT, for fault-tolerant asynchronous real-time distributed systems. The task timeliness is specified by Jensen's benefit functions and the anticipated application workload during future time intervals is given by adaptation functions. Objective is to maximize aggregate task benefit and minimize aggregate missed deadline ratio. **Rohan et al, (2012)** presents a generic checkpoint-restart mechanism based on the DMTCP checkpoint-restart package. The method is simple and can directly checkpoint the user-space QEMU virtual machines with the support of DMTCP. Therefore the generic mechanism supports only homogeneous architecture. **Chanchio et al, (2008)** design a Checkpoint Enabled Virtual Machine (CEVM) architecture to enable implicit system-level fault tolerance without modifying existing setup and to minimize the space and time overhead needed to execute software that cannot tolerate faults in HPC systems. **Arun et al, (2007)** presents an automatic proactive FT for arbitrary MPI applications to migrate an MPI task from a health-deteriorating node to a healthy node without stopping the MPI task. It uses health monitoring and load-based migration with Xen live migration mechanism. **Bogdan et al, (2013)** propose BlobCR, a dedicated checkpoint repository to take live incremental snapshots of the whole disk attached to the virtual machine (VM) instances. It aims to minimize the performance overhead of checkpointing by persisting VM disk snapshots asynchronously in the background using a low overhead call selective copy-on-write technique. **Sheng et al, (2013)** proposes VM-µCheckpoint, a lightweight software mechanism for high-frequency checkpointing and rapid recovery of virtual machines. Knowledge of fault/error latency is used to explicitly to address checkpoint corruption. **Haikun et al, (2011)** investigates design methodologies to quantitatively predict the migration performance and energy consumption and construct models for the cost prediction by using learned knowledge about the workloads at the hypervisor (VMM) level.

The list of fault tolerance research in virtualization is innumerable therefore further literature studies will not be listed since the objective to understand the nature of the presented problem simply revolves around one or two qualitative observations and therefore not comprehensive in nature. Hence even after all these efforts were made the Byzantine faults never fails to make an impact in Cloud platform and always expected to evolve to new levels. According to the performed literature study the following were the key elements observed to be considered for successful development of Byzantine fault tolerant cloud system.

### 2.3 Cloud Computing Crisis Implications

Comprehensive study on fault and the methods proposed to handle them has helped to arrive at following crisis deductions.

*Byzantine nature:* Mostly assumed faults are fail-silent faults it happens only when the faulty unit is expected to stop functioning and thus produces no bad output. However in reality mostly occurring faults are Byzantine faults. A Byzantine fault occurs when the faulty unit continues to function but produces bad output usually appears as a proper output. Dealing with Byzantine faults is extremely difficult due to its nature of disguise.

*Unsuitable Incremental nature:* Mostly derived checkpoint algorithms are incremental and therefore sequential in nature. Due to technological advancements even the single device nowadays performs parallel processing. Therefore highly parallelized platform like cloud needs parallel fault handling mechanism.

*Replication versus Redundancy:* Replication is usually the focused area for research which involves several units operating concurrently and a supervisor system to store the data in multiple locations. However with redundancy, only one unit functions while the redundant units sleep and can be invoked when the need arrives. The byzantine fault can have ill effect on both replication and redundancy; however the improved byzantine fault tolerant system can improve the redundancy. Hence the better redundancy can reduce the replication needs, falling to have a proper redundancy can demand costly, time and space consuming replication and yet may fail in case of byzantine error because the error were also getting replicated.

*Reliability and dependability* trends seem to crumble at least in hybrid cloud scenario due to high level of heterogeneity. Disparity among nodes, existence of Single point failure due to supervisor malfunction, the possibility of inconspicuous error propagation among nodes and the consequent driving of entire system into dilapidation indicates that the dependability of modern hybrid cloud is at least decreasing. This development overturns the established trend of increasing HPC dependability. Bounding the Cloud failure behaviors at node level or transient level has become increasingly more difficult due to the increase in unpredictable Byzantine trends. Strategies to mitigate the problems are yet to make an impact. Therefore it is safe to assume that the anticipated rate of byzantine failure will increase and the modes of failure will become ever more difficult to characterize.

*Unpredictable faults:* Replica damage due to intrusion has been the formidable challenge for cloud.

*Mission critical nature*: Byzantine fault is connected with performance, since most of the application nowadays have performance requirement if it fails to meet the performance requirements it is again considered as failure of cloud system. Moreover for mission critical applications delay and error is not tolerated, which is thus marked as the characteristics for byzantine fault.

## 3. Proposed Byzantine fault detection through Checksum Validation

MD5 is one of the most common hash algorithms in use today. MD5 is a 128-bit hash function. Unlike its predecessors MD2 and MD4 it is one of the fastest and secure hash functions in common use. MD5 checksum is also widely used in Cloud enabled platforms, therefore the estimation processing and storage overhead for checksum is readily available for virtual machines. The processing and storage overhead for checksum in VM platform is estimated to be just 1%. *John et al (2016)*.

### 3.1 Capacitating the intra node

MD5 is often performed at the receiving end of the TCP/IP if a node does not use TCP/IP for transmission then the checksum is considered unnecessary *John et al (2016)*. The nodes are classified based on the transient link it is connected to as follows; If the node is connected to another cloud node through TCP/IP protocol then it is expected to perform checksum so such nodes are automatically configured with MD5. Such nodes which are connected to remote node through Internet protocol is termed as **inter nodes** for the sake of modeling. However the nodes which are part of the same system is still connected to each other through a transient link which is highly virtual in nature and does not follow the TCP/IP stack and therefore doesn't involve MD5 are termed as **intra nodes**.

However the proposed work involves every node in the cloud platform to perform the checksum. Since a given data block will always results in unique checksum and does not collide with the result of another data block. Therefore when node is presented with a message and if the node doesn't produce the checksum which is relevant then the node can be spotted as erroneous are compromised. Reconstructing the original data from the checksum or performing collision analysis is usually time, cost and space constrained tasks and thus discourages the malicious nodes from manipulating the checksum results. Computing the MD5 checksum for any arbitrary data set is easy and feasible for even the resource, power, space constrained systems due to its simplicity. Moreover the MD5 checksum involves various computations as its operational procedure. The Byzantine nodes often generate genuinely looking output but are wrong due to the miscalculation induced through byzantine faults. Therefore when using the MD5 checksum to validate the node it allows a peer node to compute the checksum and propagate the message to other nodes, the replying node for a same message has to produce same output, the test the byzantine nodes may fail because of its incapability to compute incorrectly.

## 3.2 Checksum Challenge

The processing requirement of the MD5 for Cloud scenario requires sending a single message **M** by a cloud monitoring node automatically to **n** number of intra nodes and receive Message digest **{H₁, H₂, …, Hₙ}** in time **{T₁, T₂,…. Tₙ}**. The standard message M is of the size 512 bits which is the required block size for MD5 and the resultant checksum is 128 bit size, since MD5 with small hash size is considered more secure and it is easy to compute []. For **M** the supervisor or monitoring node has the pre-computed message digest **H** in time **T**. Now compare the set **A {H}** with **B {H₁, H₂, …, Hₙ}** as follows

## 3.3 Compare Checksum

If A and B are sets and every element of B is also an element of A, then: **B ⊆ A** where B is a subset of A, That is,

$$B \subseteq A \; if \; \forall x(x \in B \rightarrow x \in A)$$

This denotes that the set B has all hash values equal to the hash value of set A so the tested elements exhibits no processing errors.

If set B is not a subset of A, then **B⊄A** this denotes that one or few element of B exhibits processing error thus the difference in the hashes.

If the set B contains no elements equal to A then the null set { } and it is denoted by ∅. The null set is often the result of the intersection and it is given as follows;

$$\text{A and B are disjoint if } A \cap B = \emptyset$$

The ∅ implies that either the entire group of observed hashes is incorrect so the entire set of observed nodes is compromised. It may also imply that the supervisor itself is compromised since the hash of set A may also exhibit error.

The complement of a set A refers to elements not in A. Therefore the relative complement of A with respect to a set B, simply the difference of sets A and B gives the list of hashes which does not comply with the hash of A. In other words if there exist as set of hashes in set B which is generated by the erroneous nodes then;

$$A \setminus B = \{x : x \in A \mid x \notin B\} \Rightarrow \text{Set of erroneous hashes}$$

## 3.4 Pre-computation

However before starting the cloud services the monitoring node chooses the message M after generating the Hash A{H} and automatically send it to **n** number of required nodes and receive Message digest B **{H₁, H₂, …, Hₙ}** in time **{T₁, T₂,…. Tₙ}**. After making sure the chosen set $B \subseteq A \; if \; \forall x(x \in B \rightarrow x \in A)$ it allots the nodes to the customer and records the response time (processing time + transit time) as set **X {T₁, T₂,…. Tₙ}**.

## 3.5 Operating procedures

Providing cloud service involves establishing the SLA with the customers, it involves various QoS metrics such as among those the response time (RT) is chosen because it is simple to observe and ominously present in all the SLA agreement therefore associated with every cloud nodes. The set of nodes which is obliged by the Cloud service provider is observed by the destined node or supervisor.

**Algorithm 1: Byzantine Error Detection**

**for each** *Virtual node (v)*
   **if** *response time of* $v_i \geq$ **QoS** *response time*
                  //**Where $v_i$ is any node from the set of operating nodes {$v_1, v_2….v_n$}**
   **then**
      **call** checkpoint
      **call** *checksum challenge()*
   **else**
      **continue** monitoring
   **end if**
**end for**

According to algorithm 1 for a data block if the response time for any node exceeds its QoS requirement then the node is check pointed and the following algorithm is called.

**Algorithm 2: Checksum Challenge ()**

**Challenge $n_i$** *with M*
    **if** *H' ≠ H*
       //**denotes Byzantine error**
   **then**
        **Shutdown $v_i$**
        **Start** new node as $v_i$
   **else**
      **Call** C*ompare delay variation ()*
   **end if**

Checksum Challenge is nothing but a collision detection, MD5 collision detection is easy and fast due to the avalanche effect []. It means even a small change in the message will result in a mostly different hash overwhelmingly. Due to the avalanche effect even if a single bit in the input is changed slightly the output hash changes drastically. The algorithm 2 present the node or nodes under observation with pre-chosen message **M** if it produces result which is deviant from the **H** then it exhibits checksum error either caused by the virtual link error or by the node error which denote the presence of *Byzantine error*. If a byzantine fault is detected then the algorithm shuts the node and starts the new VM machine. If no checksum error is detected then the partial set **Y{ $T_1, T_2,…. T_i$ }** is compared with set **X {$T_1, T_2,…. T_n$}**

**Algorithm 3: Compare delay variation ()**

**for each $v_i$**
    **Select** *$T_i$ in Y*
    **Copy** *corresponding $T_i$ in X*

```
        if T_i in Y < T_i in X
            null fault              //delay variation is minimal
            Call checkpoint optimization ()
        elseif T_i in Y = T_i in X
            null fault              //delay variation is zero
            Call checkpoint optimization ()
        elseif T_i in Y ≥ upper bound in X
            Calculate  supremum
                if  T_i in Y ≥ supremum in X
                    Shutdown v_i
                    Start new node as v_i
                else
                    Call checkpoint optimization ()
                end if
            end if
    end for
```

### 3.6  Delay variation checkpoint

Given a function **f** with domain **X** and a partially ordered set **(Y, ≤)** as codomain, an element **y** of **Y** is an upper bound of **f** if **y ≥ f(x)** for each **x** in **X**. if the comparison holds true for at least one value of x. Then it indicates the delay variation experienced is significant enough 'high' therefore checkpoint is set and replication is made. Now choose a constant value **z** such that **z<x** and the upper bound is calculated as **(z+ x)** where **x ≥ X {T₁, T₂,…. Tₙ}**. The upper bound is now set as *supremum* since no smaller value is an upper bound.  Now if **y ≥ f (z + x)** then it is "extreme" either it denotes Byzantine failure or performance failure therefore the node is shutdown after automatic transfer of workload at previous checkpoint.

### 3.7  Discrete state modeling

Cloud system is considered a discrete system because the assumed state variable as delay variation changes at a discrete set of points in time. Moreover the cloud system as a discrete system has only countable number of states. The Cloud system is considered a discrete model to determine the virtual system behavior at varying time intervals. These conditions are represented by the occurrence of events that cause specific component mode transition from *fail safe* to *fail-stop or byzantine prone*. During the observation both the discrete component modes and the set of system state variables need to be tracked. Accordingly, the overall system state at time **t** is described by

$$X(t) = \Theta\ (c(t), d(t))$$

Where **X(t)** is the overall system state**, c(t) = [c₁,c₂,c₃, ... .cₙ]** is a vector of discrete component modes computed with delay variation (**Y**) for each component **c = 1, ... .n** (n: number of nodes in the virtual system). Assumes a discrete mode from its own set of **M** modes where **c_i= (c_{i1}, c_{i2}, c_{i3}, ... . c_{iM}},** and **d(t}= (d₁,d₂,d₃, ... . ,d_K)** are the vector of system state variables. During conceptual design, the system state variables are not known quantitatively. Therefore, these continuous variables are assume as discrete since the system only involves finite number of states and the set of qualitative values are assumed. The vector **d(t)** with variable as checksum(¢) then defines these qualitative values for each state variable **v$_i$** from a set of **P** possible values **d$_i$**= (**d$_{i1}$,d$_{i2}$,d$_{i3}$, ... . ,d$_{iP}$**). A delay variation variable takes on values from the set of **P {low, normal, high, and extreme}**. Similarly, a check sum error variable takes on values from the set of **P {no error, error}**.

The proposed system involves a lightweight state model since maintaining state information is essential for virtual systems to reverse the fault with the previously checkpointed data. The proposed model is designed to be simple since it involves processing small checksum and comparative delay variation monitoring. Therefore it is extendable because it allows additional variables if needed, and it is flexible since it creates a possibility to optimize intervals at which the checkpoint is placed which is usually fixed and regular therefore incurs hefty cost.

## 4. Computation and Optimization

Optimizing the proposed discrete model for a possibility to perform quick fault detection in lesser number of steps is analyzed using various state transition computations. Moreover improving Byzantine fault detection using fine-tuning the Delay variation with the Checksum is also analyzed. Setting the interval for virtual system state is fixed for regular interval and every node in the cloud system is investigated at the set interval for fault tolerance. However through optimization and analysis we try to optimize the interval and the number of nodes involved for investigation.

### 4.1 State Transition with Delay Variation

However for investigative analysis each variable involved in the algorithm has observed using state transition diagram individually. Initially the state transition has been obtained for delay variation (Ɣ) independently. The state transition diagram and table were obtained for virtual node states **{fail-safe, Byzantine, fail-stop}** with respective state variables **{S$_0$, S$_1$, S$_2$}**. In all the cases S$_2$ is considered as acceptor, since the transitions may lead up to the S$_2$ when the transition reaches the S2 the checkpoint activation follows the shutting down of respective node. The Input as delay variation with the possible set **P {low, normal, high, extreme}** is represented as **{00, 01, 10, 11}**.

**Table 1.** State transition with delay variation

| Present State | Next State | | | | Output |
|---|---|---|---|---|---|
| | 00 | 01 | 10 | 11 | |
| S$_0$ | S$_0$ | S$_0$ | S$_1$ | S$_2$ | 0 |
| S$_1$ | S$_0$ | S$_0$ | S$_1$ | S$_2$ | 0 |

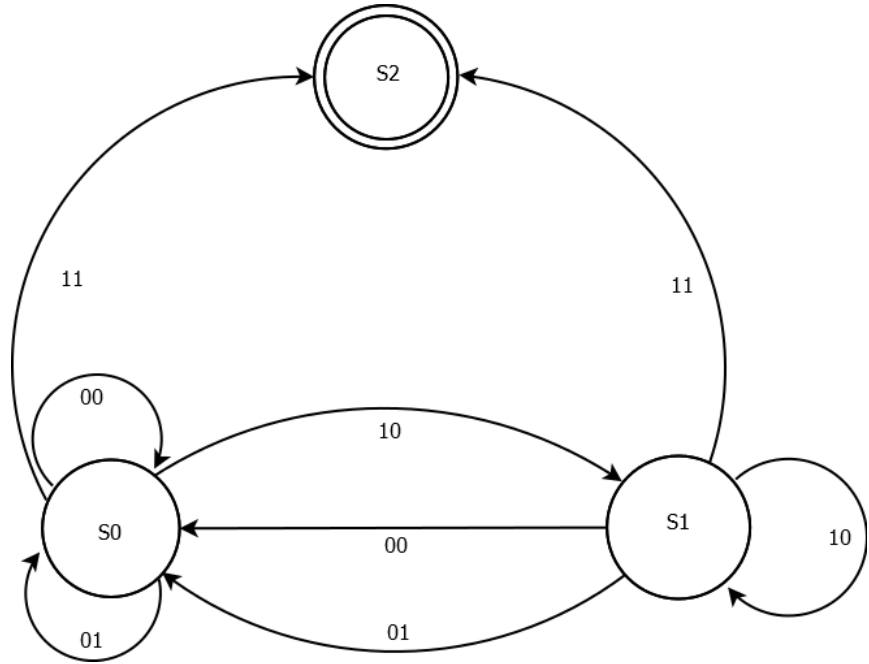

**Figure 1.** *State transition diagram with delay variation*

Byzantine fault Generalization considering the delay variation (Ɣ) alone is limited. Since the case when delay variation is high Ɣ = **10** implies that the delay variation is not *supremum* therefore not decisive in nature, which makes the Output indecisive as well. Therefore to mark it as byzantine failure and checkpointing using delay variation alone may result in incorrect detection.

### 4.2   State Transition with Checksum

The state transition for Checksum (¢) has been investigated separately as follows. The state transition diagram and table were obtained for virtual node states **{fail-safe, Byzantine, fail-stop}** with respective state variables **{$S_0$, $S_1$, $S_2$}.** In all the cases when the transition reaches the S2 then the corresponding node is shutdown. The Input as checksum involves the possibility set **P {no error, error}** represented by corresponding binary inputs **{0, 1}.**

**Table 2.** State transition with decisive Checksum

| Present State | Next State | | Output |
|---|---|---|---|
| | 0 | 1 | |
| $S_0$ | $S_0$ | $S_2$ | 1 |
| $S_1$ | $S_0$ | $S_2$ | 1 |

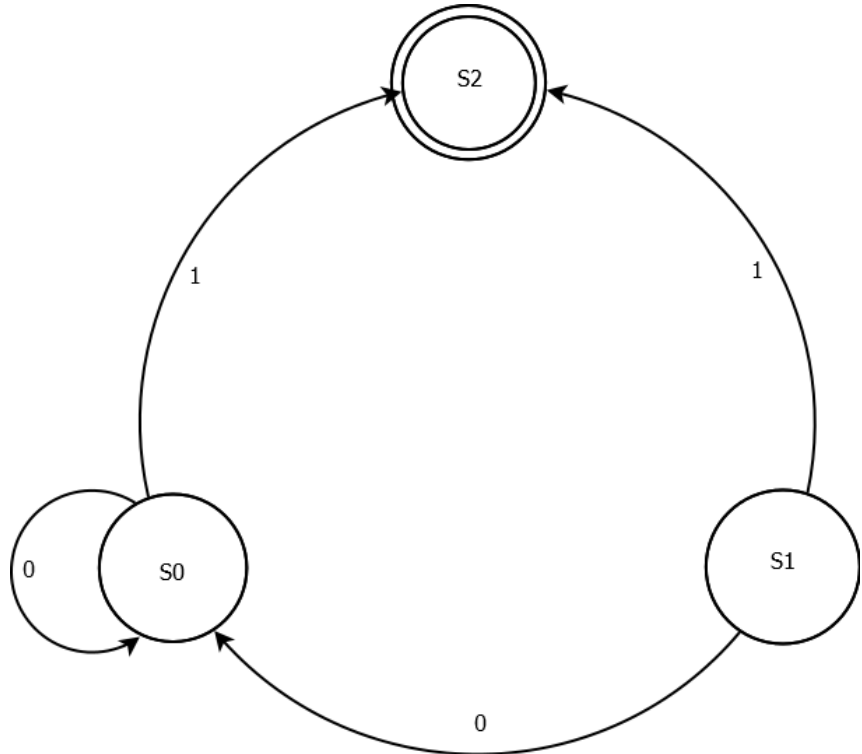

**Figure 2.** *State transition diagram for Checksum*

Byzantine fault Generalization considering the Checksum (¢) alone is more decisive in nature. Moreover it seems have eliminated the need for state transition $S_0 \rightarrow S_1,$ which implies that the checksum implementation can help to validate the virtual nodes to a greater extent. However MD5 can help to generalize the Byzantine error which happens due to faulty nodes and the link errors. In both the cases the data for hashing becomes invalid to a greater extent due to the avalanche effect. However in case if the node is compromised due to the presence of malicious element as a worst case scenario if the malicious component is equipped to allow the checksum operation intact it still may produce the relevant checksum. In such cases, therefore the decisive nature of MD5 checksum may not be simply enough. Therefore the checksum which helps to detect the byzantine faults may not be helpful to detect all the classes of evasive malicious compromise.

**4.3 Checkpoint optimization with simplified State Transition**

Optimizing the Checkpoint/Restart is often considered tricky since it is expected to operate within the affordable cost []. The state periodically and the number of tasks considered for persistent storage induces a runtime overhead and occupies more storage space and thus incurs the cost[]. Therefore the tradeoff is to control and keep the performance overhead involved with the checkpoint mechanism at acceptable levels even when the Cloud deployment grows in size and complexity [].

**Table 3.** State transition with decisive Checkpoint and simplified delay variation

| Present State | Next State | | | | Output |
|---|---|---|---|---|---|
| | 00 | 01 | 10 | 11 | |
| $S_0$ | $S_1$ | $S_2$ | $S_2$ | $S_2$ | 1 |
| $S_1$ | $S_1$ | $S_2$ | $S_2$ | $S_2$ | 1 |

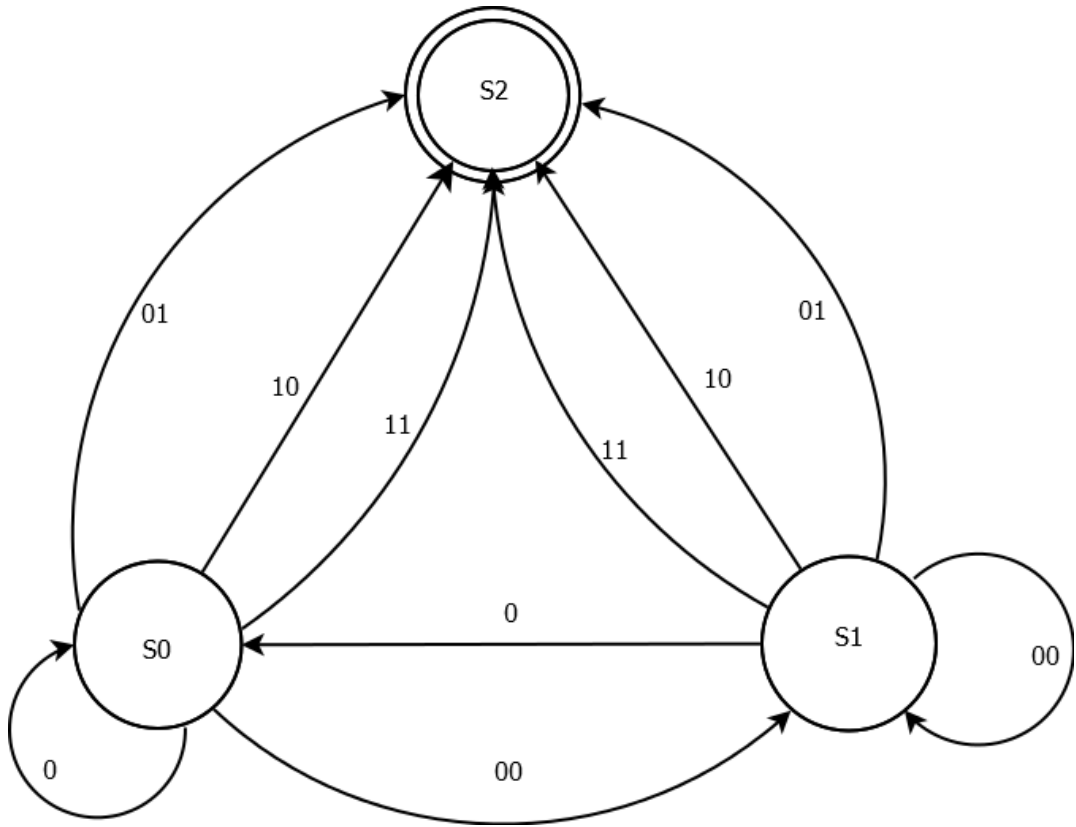

**Figure 3.** *State transition* with decisive Checkpoint and simplified delay variation

The table 3 is plotted with delay variation **P {high, extreme}** since those are the delay that intimate that the observed virtual node may be transpiring to erroneous state or already erroneous. The values are assigned as **high = 0** and **extreme = 1.** Similarly, the checksum variable takes on values from the set of **P {no error, error}** as **no error = 0** and **error = 1**. However if **no error** in checksum and if the delay variation is not high or **extreme** then the input for both Ɣ¢ =0 this can happen only in state S1. This implies that after exhibiting high Ɣ in previous state which caused the observing node $n_i$ to transition from $S_0 \rightarrow S_1$, it has recovered from the setback and for the current state the transition occur from $S_1 \rightarrow S_0$. Therefore in the current state observation the Ɣ seems to be either **low** or **normal** and thus makes the Ɣ input missing. This helps to generalize the nodes which though shows no checksum error but shows slightly increased delay over consecutive state observations.

**Algorithm 4:** *Checkpoint Optimization ()*

  *initialize x = j*                      //where j is the pre-set initial state monitoring interval
  **for each** *vi in S0,*
    **if** $Y\mathcal{c} = \{0\}$
       **Assign** *interval j = x+j*   //reducing the processing overhead by increasing interval
       **Call** *Compare delay variation ()*
    **endif**
  **endfor**

  *initialize q = 0*   //a variable to monitor the staying node for successive intervals
  **for each** *vi in S1;* $Y\mathcal{c} == \{00\};$ *q+1*
       **Set** j = x
       **Call** *Compare delay variation ()*
       **Call** *Checksum Challenge ()*
    **if** *q ==3*
      **Shutdown** $v_i$
      **Start** new node as $v_i$
    **endif**
  **end for**

For the nodes that remain in $S_0$ the next interval for monitoring can be increased from interval **j** to **2j** and after **2j** it still remains in S0 the interval can be further shifted to **3j** etc. Nodes in state $S_1$ alone needs frequent monitoring for every interval **j**, If the node in $S_1$ transitioned to state $S_0$ then the successive interval can be increased as long as it stays in $S_0$. However if it transitioned back to $S_1$ then the interval is reduced to initial value. Moreover for $i^{th}$ node if it stays at high for **three** successive intervals in state $S_1$ then it could be suspended for evaluation. This way the performance improvement can be achieved even for the normal case.

    Therefore the crucial overhead reduction is achieved through minimization of the interruption time and through the signification reduction of the number of Checkpoints implemented not only for the erroneous case but also for normal case. Moreover restart can be promptly initialized with previously saved checkpoint with minimal overhead. Thus the considered discrete model is configured through various state transition computations and optimization.

### 4.4 Byzantine Fault Generalization

A Cloud system is said to be k-fault tolerant if it can withstand k faults. If the virtual node fail silently then it is sufficient to have **k+1** component to achieve k fault tolerance. Since even if the **k** components fail the cloud system still be able to work with k replaced nodes and still has **1** redundant node for replacement. However the success of Fault tolerance mechanisms varies when it comes to tolerating faults. A system can only tolerate crash faults with **k+1** replication that to with passive replication. However solutions to the Byzantine fault tolerance usually are complicated therefore to tolerate byzantine faults it requires active replication system which uses

**3k +1** redundant nodes. However using the checksum for fault detection the byzantine fault tolerance requirement can be reduced to just **K+1** replica since it is capable to generalize the Byzantine fault as crash faults. Therefore active replication becomes possible with online delay variation analysis.

## 5. Result Analysis

Evaluating the set of proposed algorithms to understand the real-time implications requires simulation and real world data analysis which is presented as follows.

### 5.1 Experimental Setup and Simulation

The CloudSim is a development toolkit for discrete event simulation of Cloud computing *Rodrigo et al, (2010)*. CloudSim supports modeling of Cloud data centers and its simulation with different hardware configurations. In addition, this simulation toolkit helps in modeling a range of virtual machines having independent tasks, design, different resources and VM provisioning.

Using modified fault generators in FailureGenerator Class the byzantine fault inducer has been created. FaultTolerantScheduling module is extended to test the byzantine faults but it fails detect the byzantine faults because it does not drive the node to failure. Moreover FailureGenerator which doubles the size of distribution samples each time but only to maxFailureSizeExtension parameter. This is to limit the failure rate from reaching a point where it cannot be handled. In our case since the error rate is required for all the monitoring interval maxFailureSizeExtension has been set to maximum. Furthermore the cloud fault tolerance module is extended to monitoring the faults since it is not equipped to handle byzantine faults it exhibit limited fault generalization capabilities. Therefore by extending FileAttribute class the checksum attribute is set to MD5. This class also has the collision detection i.e. *if (checksum < 0) {return false;}* inherently build in it, using this better error generalization capability has been constructed.

The performance of MD5 is comprehensively studied from various perspectives due to its vast usage. Therefore we don't get into those details. However the reason for widespread usage of MD5 in Cloud computing is evident in its speed. Among observed checksums MD5 is observed to perform computations very fast even on the 32-bit processors and less CPU intensive. Furthermore MD5 is observed to be more secure in case of smaller messages than the larger messages. It is thus considered as the most appropriate algorithm with essential qualities for observing the Virtual nodes behavior critically.

The Failure tolerant system of CloudSim is designed to detect the node crashes but not the errors. Therefore the induced Byzantine error does not derive the nodes to crash therefore evade the fault tolerant system. However the checksum collision observed is able to detect all the erroneous nodes. It is visible in the following graph since the error generated are based on the exponential increase the error generalization also follows the exponential increase pattern so it is obviously evident.

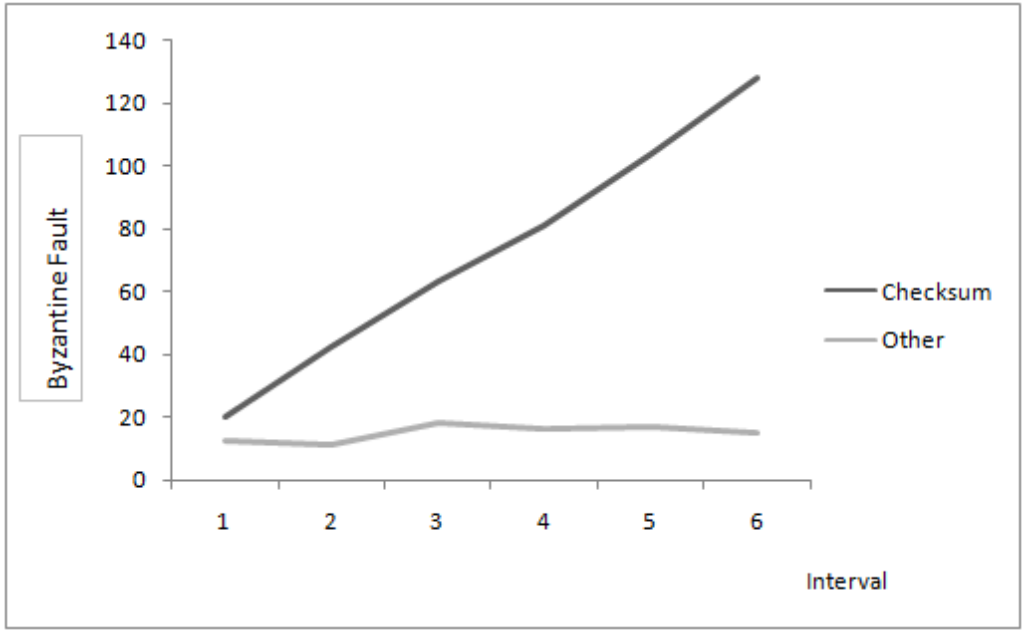

**Figure 4.** *Byzantine Fault detection with and without MD5 Checksum*

Figure 4 shows the suitability of using the checksum algorithm to detect byzantine faults as promising. However the challenge is to detect the pure byzantine fault which is caused due to the concealed security breaches. Until now the CloudSim doesn't provide support for such scenarios.

### 5.2  Optimization Evaluation

As a result, the Google cloud dataset has been obtained from github.com for further analysis. In the data set User and job names are hashed and provided as base 64-encoded strings that can be tested for equality. Additionally all jobs and tasks have a scheduling class that represents the delay sensitive requirements, this validates the initial claim in choosing the response time as observable variable for performance overhead optimization. Performance measurements are taken for every pre-set interval which is '1' second. Conversely the response time is reported in microseconds. The system load sometimes seems to prevent the resource metrics from being sampled at the desired interval, so one or more intervals may span more than a 1 second. The provided sampling rate as the ratio between the number of expected samples to the number of observed samples helps to detect the interval variation. This is ironic to the rate reduction interval proposed in the Checkpoint Optimization () algorithm. Therefore validating the proposed algorithm through deriving possibilities to increasing the interval becomes our main objective in sampling the dataset. Moreover checksum has been computed by the virtual nodes as just another task among various other tasks. This leaves a possibility for malicious element to configure the compromised node in such a way to allow processing the checksum intact while inducing miniscule errors in other tasks.

The checksum anomaly and delay variation has been observed throughout the data set with sample space of 500 nodes. The 539 hash collision has been detected as the result of analyzing the entire dataset. The delay variation and the hash collision percentage for the obtained result has been computed and plotted as following scatter diagram.

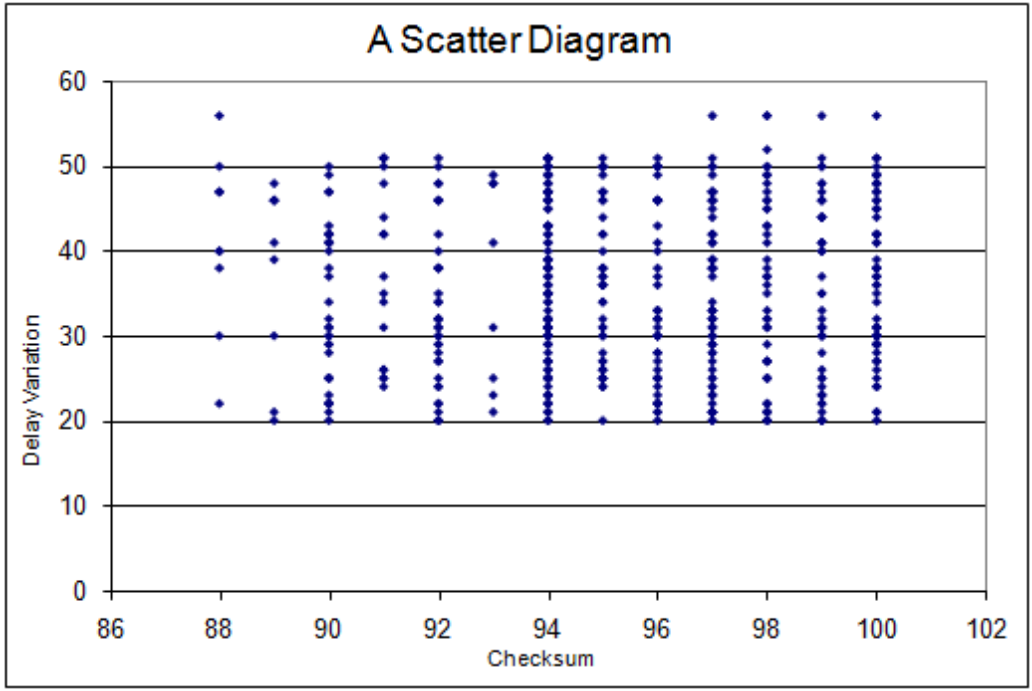

**Figure 5.** *MD5 and delay variation based fault detection possibilities (%)*

According to Figure 5 for all the cases, the percentage of variation in hash due to the modification of input string remains over 90 percent consistently, which is attributed to avalanche effect. This can help in generalizing the Byzantine errors quickly because the checksum comparison algorithm can detect the Byzantine error with just one checksum variation. Often found in immediate places from the left of the string, thus eliminate the need for comparing the entire string. Moreover in all those cases the delay variation becomes evident, where 20% to 30% indicates that the $Y = 1$ and the higher percentage indicates $Y = 0$. This shows that the chosen parameters are good fit for byzantine error/fault generalization.

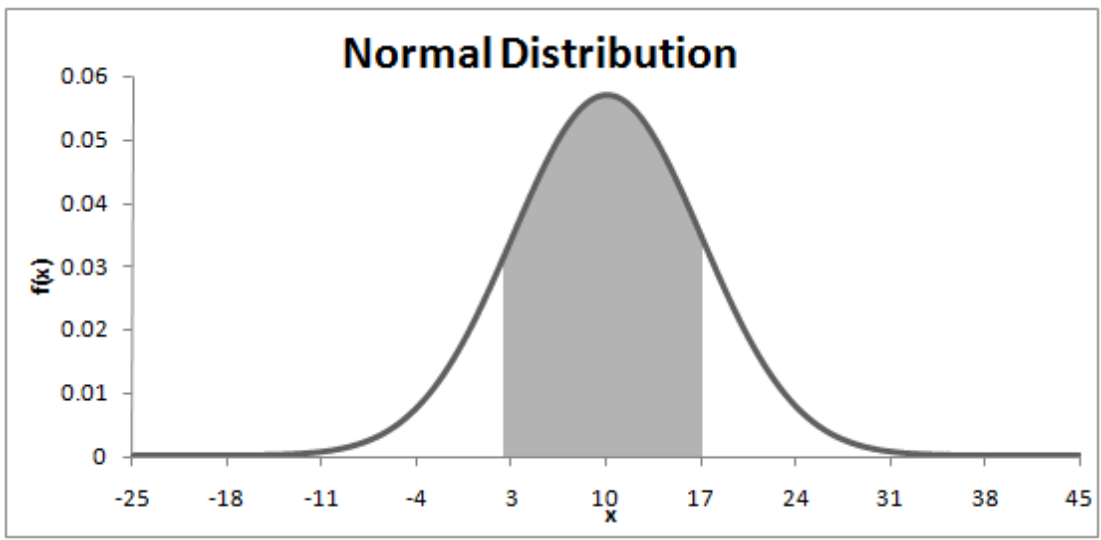

**Figure 6.** *Fault generalization percentage for delay variation*

According to figure 6 the overall delay variation (Y) including both {0,1} possibilities for all the instances in the dataset is <17% . However within the overall 17% delay the Y = 1 is 38% and the Y = 0 is 62%.. However through excluding the Y involved in checksum error instances i.e. 64% the percentage of both delay variation become 10.88%. This indicates that, for certain instances even though the ¢ = 0 the delay variation still may remain at extreme. Therefore it could very well indicate the presence of concealed malicious elements.

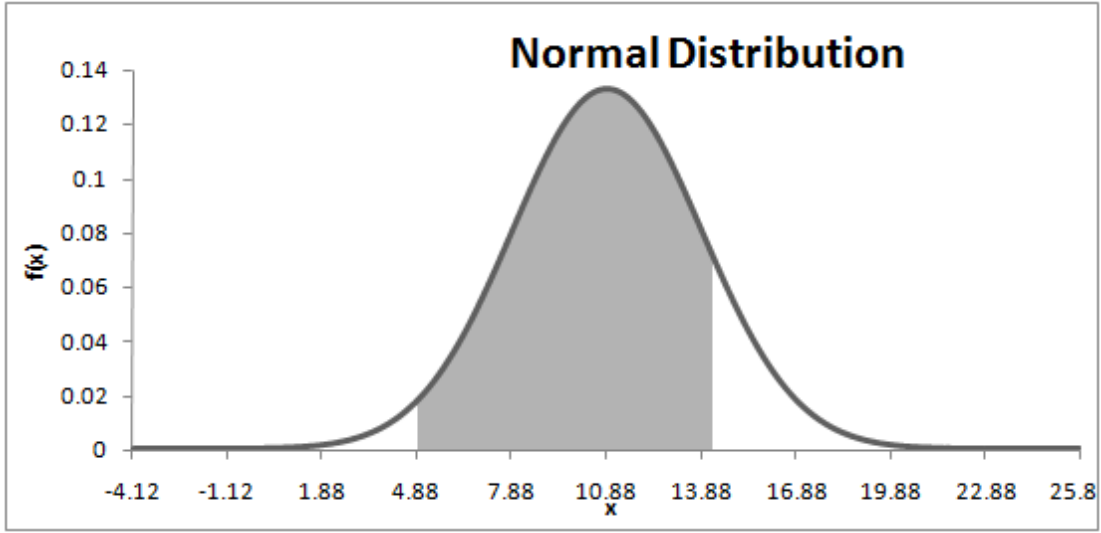

**Figure 7.** Optimized *Observable range with Y = 0*

In the 10.88 % excluding the delay variation Y = 1 since the node exhibiting it can be transitioned to state S2. Observable range for delay variation just comes to Y = 0 which is obtained with the normal distribution as between 4.88% to 14%. Therefore choosing the delay variation and checksum can minimize the processing overhead for regular intervals to 14% otherwise as 100% for every interval.

## 6. Conclusion and Future Work

A Byzantine fault is unpredictable and undetectable since the faulty unit continues to function but produces incorrect outputs often appears as correct output. However Fault tolerance can only be achieved in cloud systems if only the system have the capability to detect the byzantine errors before it propagate throughout the system. Therefore the proposed work involves every node in the cloud platform including intra nodes to perform the checksum. Since a given data block will always results in a unique hash even a single error due to byzantine fault can be detected. Using those variables a lightweight discrete state is modeled since maintaining state information is essential for virtual systems to reverse the fault with the previously checkpointed data. The considered discrete model is configured through various state transition computations and optimization. The crucial overhead reduction is achieved through minimization of the interruption time and through the signification reduction of the number of Checkpoints implemented not only for the erroneous case but also for normal case. Moreover restart can be promptly initialized with previously saved checkpoint with minimal overhead. The model is designed to be simple since it involves processing small checksum and comparative delay variation monitoring. Therefore it is extendable because it allows additional variables if needed, and it is flexible since it creates a possibility to optimize intervals at which the checkpoint is placed which is usually fixed and regular therefore incurs hefty cost.

The proposed model has been simulated using CloudSim, the results show that the suitability of using the checksum algorithm to detect byzantine faults as promising, since the checksum collision observed is able to detect all the erroneous nodes. However the challenge is to detect the pure byzantine fault which is caused due to the concealed security breaches. As a result, the Google cloud dataset has been obtained from github.com for further analysis. While sampling the dataset since the checksum has been computed by the virtual nodes as just another task among various other tasks, the possibility for malicious element to configure the compromised node in such a way to allow processing the checksum intact while inducing miniscule errors becomes evident. According to the result analysis, through involving delay variation considerable percentage of errors due to the concealed malicious elements has been generalized. The solutions to the Byzantine fault tolerance usually are complicated therefore to tolerate byzantine faults it requires active replication system which uses **3k +1** redundant nodes. However using the proposed model which involves checksum and delay variation for fault detection the byzantine fault tolerance requirement can be reduced to just **K+1** replica. Hence the attempted modeling with decisive checksum and supportive delay variation has helped us to effectively optimize the checkpoint interval and number of nodes required for observation for byzantine fault as well as for deterministic cases with malicious elements.

However to capacitate it for more malicious fault generalization the proposed model will be extended for multivariate based vector inference in our future work. Moreover the possibility for parallel implementation in intra nodes to develop an active correlation analysis and the means to tolerate the single point failure in supervisor controlled cloud will be investigated in the future work.

**References**

[1]. J. P. Martin and L. Alvisi, "Fast Byzantine Consensus", IEEE Transactions on Dependable and Secure Computing, Vol. 3, Iss. 3, Aug. 2006.
[2]. Christian Cachin, Idit Keidar and Alexander Shraer, "Trusting the cloud", ACM SIGACT, Vol. 40, Iss. 2, June 2009 pp. 81-86.
[3]. Kevin Driscoll, Brendan Hall, Håkan Sivencrona and Phil Zumsteg, "Byzantine Fault Tolerance from Theory to Reality", Springer International Conference on Computer Safety, Reliability, and Security (SAFECOMP), pp 235-248, 2003.
[4]. Miguel Castro and Barbara Liskov, "Practical Byzantine Fault Tolerance", USENIX Proceedings of the Third Symposium on Operating Systems Design and Implementation, pp. 173–186, Feb. 1999.
[5]. Thomas C. Bressoud and Fred B. Schneider, "Hypervisor-based Fault-tolerance", Proceedings of the fifteenth ACM symposium on Operating systems principles SOSP '95, pp. 1-11, Dec. 1995.
[6]. Yilei Zhang, Zibin Zheng and Michael R. Lyu, "BFTCloud: A Byzantine Fault Tolerance Framework for Voluntary-Resource Cloud Computing", IEEE 4th International Conference on Cloud Computing, pp. 444-451, July 2011.

[7]. Jun Zhu, Wei Dong, Zhefu Jiang, Xiaogang Shi, Zhen Xiao and Xiaoming Li, "Improving the performance of hypervisor-based fault tolerance", IEEE International Symposium on Parallel & Distributed Processing (IPDPS), pp. 44-51, April 2010.

[8]. Bhaskar Prasad Rimal, Eunmi Choi and Ian Lumb, "A taxonomy and survey of cloud computing systems", IEEE Fifth International Joint Conference on INC, IMS and IDC (NCM '09), Aug. 2009.

[9]. Rudiger Kapitza, Johannes Behl, Christian Cachin, Tobias Distler, Simon Kuhnle, Seyed Vahid Mohammadi, Wolfgang Schr, oder-Preikschat and Klaus Stengel, "CheapBFT: resource-efficient byzantine fault tolerance", Proceedings of the 7th ACM European conference on Computer Systems EuroSys '12, pp. 295-308, April 2012.

[10]. Pedro Costa, Marcelo Pasin, Alysson N. Bessani and Miguel Correia, "Byzantine Fault-Tolerant MapReduce: Faults are Not Just Crashes", IEEE Third International Conference on Cloud Computing Technology and Science (CloudCom), Dec. 2011.

[11]. Pierre-Louis Aublin, Sonia Ben Mokhtar and Vivien Quema, "RBFT: Redundant Byzantine Fault Tolerance", IEEE 33rd International Conference on Distributed Computing Systems (ICDCS), July 2013.

[12]. Ravi Jhawar and, Vincenzo Piuri, "Fault Tolerance and Resilience in Cloud Computing Environments", 2nd Edition, J. Vacca (ed), Morgan Kauffmann, Computer and Information Security Handbook, 2013.

[13]. Dominic Lucchetti, Steven K. Reinhardt and Peter M. Chen, "ExtraVirt: detecting and recovering from transient processor faults", Proceedings of the twentieth ACM symposium on Operating systems principles SOSP '05, pp 1-8, 2005.

[14]. Chunye Gong, Jie Liu, Qiang Zhang, Haitao Chen and Zhenghu Gong, "The Characteristics of Cloud Computing", 39th IEEE International Conference on Parallel Processing Workshops (ICPPW), pp. 275-279, Sept. 2010.

[15]. Rafael R. Obelheiro, Alysson Neves Bessani, Lau Cheuk Lung and Miguel Correia, "How Practical are Intrusion-Tolerant Distributed Systems?", Department of Informatics, University of Lisbon, Sep-2006.

[16]. Ifeanyi P. Egwutuoha, David Levy, Bran Selic and Shiping Chen, "A survey of fault tolerance mechanisms and checkpoint/restart implementations for high performance computing systems", Springer The Journal of Supercomputing, Vol.65, Feb. 2013, pp. 1302–1326.

[17]. Stephen L. Scott, Geoffroy Vallée, Thomas Naughton, Anand Tikotekar, Christian Engelmann and Hong Ong, "System-level virtualization research at Oak Ridge National Laboratory", Elsevier Future Generation Computer Systems, Vol. 26, 2010, pp.304-307.

[18]. Binoy Ravindran, Peng Li, and Tamir Hegazy, "Proactive resource allocation for asynchronous real-time distributed systems in the presence of processor failures", Elsevier Journal of Parallel and Distributed Computing, Vol.63, 2003, 1219–1242.

[19]. Rohan Garg, Komal Sodha and Gene Cooperman, "A Generic Checkpoint-Restart Mechanism for Virtual Machines", arXiv:1212.1787 [cs.OS], Dec. 2012.

[20]. K. Chanchio, C. Leangsuksun, H. Ong, V. Ratanasamoot and A. Shafi, "An Efficient Virtual Machine Checkpointing Mechanism for Hypervisor-based HPC Systems", Proceeding of the High Availability and Performance Computing Workshop (HAPCW), 2008.

[21]. Manav Vasavada, Frank Mueller, Paul H. Hargrove and Eric Roman, "Comparing different approaches for Incremental Checkpointing: The Showdown", Proceedings of theLinux Symposium, June 2011.
[22]. Arun Babu Nagarajan, Frank Mueller, Christian Engelmann and Stephen L. Scott, "Proactive Fault Tolerance for HPC with Xen Virtualization", ACM Proceedings of the 21st annual international conference on Supercomputing ICS '07, pp. 23-32, June 2007.
[23]. Bogdan Nicolae and Franck Cappello, "BlobCR: Virtual disk based checkpoint-restart for HPC applications on IaaS clouds", Journal of Parallel Distribed Computing, vol. 73, 2013, pp. 698–711.
[24]. Sheng Di1, Yves Robert, Frédéric Vivien, Derrick Kondo, Cho-Li Wang and Franck Cappello, "Optimization of Cloud Task Processing with Checkpoint-Restart Mechanism", ACM Supercomputing, Nov. 2013.
[25]. John Nicholson, Jase McCarty and Jeff Hunter, "VMware Virtual SAN™ 6.2 Space Efficiency Technologies", Technical White Paper, March 2016, pp.1-15.
[26]. Haikun Liu, Hai Jin, Cheng-Zhong Xu, and Xiaofei Liao, "Performance and energy modeling for live migration of virtual machines", Springer Cluster Computing, vol. 16, Iss.2, pp.249–264, June 2013.
[27]. Rodrigo N. Calheiros, Rajiv Ranjan, Anton Beloglazov, C´esar A. F. De Rose and Rajkumar Buyya1, "CloudSim: a toolkit for modeling and simulation of cloud computing environments and evaluation of resource provisioning algorithms", Wiley's Software Practice and Experience, Aug. 2010, pp.23–50.